\def\CHAP{}
\newcount\eqnumber

\def\eq(#1){
    \ifx\DRAFT\undefined\def\DRAFT{0}\fi	
    \global\advance\eqnumber by 1%
    \expandafter\xdef\csname !#1\endcsname{\the\eqnumber}%
    \ifnum\number\DRAFT>0%
	\setbox0=\hbox{#1}%
	\wd0=0pt%
	\eqno({\offinterlineskip
	  \vtop{\hbox{\the\eqnumber}\vskip1.5pt\box0}})%
    \else%
	\eqno(\CHAP\the\eqnumber)%
    \fi%
}
\def\(#1){(\CHAP\csname !#1\endcsname)}

\font\ss = cmssbx10
\font\title = cmbx10 scaled 1440        
\def\ct#1{\centerline{\title #1}}
\magnification=\magstep1
\baselineskip=13pt
\parskip = 2pt
\def\>{\rangle}
\def\<{\langle}
\def\k#1{|#1\>}
\def\Xc{\hbox{\ss X}}
\def\Zc{\hbox{\ss Z}}
\def\Pc{\hbox{\ss P}} 
\def\Uc{\hbox{\ss U}} 
\def\Mc{\hbox{\ss M}}
\def\Ac{\hbox{\ss A}}
\def\Hc{\hbox{\ss H}}
\def\fr#1/#2{{\textstyle{#1\over#2}}} 
\def\lp{\bigl(}
\def\rp{\bigr)}
\def\h{\fr1/2}
\def\oc{\hbox{\ss 1}}
\def\al{\alpha}
\def\be{\beta}

\input epsf
\ct{In praise of measurement}
\vskip 10pt
\centerline{N. David Mermin}

\centerline{Laboratory of Atomic and Solid State Physics}

\centerline{Cornell University, Ithaca, NY 14853-2501}

\vskip 20pt

{\narrower

{\bf Abstract.} The role of measurement in quantum computation is
examined in the light of John Bell's critique of the how the term
``measurement'' is used in quantum mechanics.  I argue that within the
field of quantum computer science the concept of measurement is
precisely defined, unproblematic, amd forms the foundation of the
entire subject.

}

\vskip 20pt

{\narrower \narrower

\noindent {\sl Here are some words which $\ldots$ have no place in a {\it
formulation\/} with any pretension to physical precision:
{\it system, apparatus, environment, microscopic, macroscopic,
reversible, irreversible, observable, information, measurement.}
On this list of bad words the worst of all is
``measurement''. $\ldots\,\,$What exactly qualifies some physical
systems to play the role of ``measurer''? $\ldots$ The word has had
such a damaging effect on the discussion, that I think it should now
be banned altogether in quantum mechanics.}  \vskip 5pt 

\hskip 150pt --- J.~S.~Bell [1]

\vskip 15pt

\noindent {\sl In our description of nature the purpose is not to
disclose the real essence of the phenomena but only to track down, so
far as it is possible, relations between the manifold aspects of our
experience.}

\hskip 150pt --- Niels Bohr [2] 

}

\vskip 35pt
In his his elegant tirade against measurement John Bell declared that
we lack ``an exact formulation of some serious part of quantum
mechanics.''  He explained that by ``exact'' he meant ``fully
formulated in mathematical terms, with nothing left to the discretion
of the theoretical physicist.''  And by ``serious'' he meant that
``some substantial fragment of physics should be covered'' and that
``$\,$`apparatus' should not be separated off from the rest of the
world into black boxes.''

It's sad that Bell's early death deprived him of the pleasure of
experiencing the quantum-computation revolution, and a misfortune for
physics that it deprived us of the critical insights he surely would
have had into its implications for quantum foundations.  I don't think
Bell would have denied that the theory of quantum computation is as
exact as any branch of applied mathematics can be said to be, with
nothing left to the discretion of the user beyond the choice of
problem to which to apply it, and his or her ingenuity in devising
quantum algorithms.

Whether he would have viewed quantum computation as a ``serious part
of quantum mechanics'' is less clear.  I would maintain that any
fragment of physics, large enough to be applied to the efficient
factoring of enormous integers, has to be viewed as substantial.  The
U.S.~National Security Agency surely does.  But there is no denying
that the computational apparatus --- the actual machinery of the
computer --- is separated off into black boxes from the Qbits on which
it acts.  [I digress to commend the term ``Qbit'', as a highly
convenient abbreviation for ``qubit'', especially in contexts where
one also talks about Cbits, as the physical carriers of a bit of
information in a classical computer.]  But that machinery is also,
unproblematically, separated off into black boxes from the Cbits on
which it acts in classical computer science. 

I'm not sure Bell would have found any of the remarks that follow
compelling, or even suggestive.  Nevertheless, I have found it
illuminating to reexamine the role of measurement in quantum
computation from the perspective of his critique, and I offer such
a reexamination as a 60th birthday present to Anton Zeilinger, who I
hope will take a more sympathetic view of it than John Bell might have
done, if we had had the good fortune to have him with us here today.

The first striking change in how quantum mechanics appears, when
viewed through its serious subset of quantum computation, is that the
continuous time evolution of ordinary quantum mechanics is replaced by
the actions of a collection of discrete gates.  Some important
quantum-computational gates are shown in Figure 1, together with the
gate of special interest to us, the 1-Qbit measurement gate.  All of
the gates in Figure 1, including the measurement gate, alter the state
associated with the incoming Qbits in a well-defined, generally
discontinuous manner, which is precisely defined by that state, with
nothing left to the discretion of the theoretical physicist.  Within
this framework, the action of the measurement gate differs from the
other gates, all of which are unitary, in several ways.

\midinsert
\epsfxsize=6truein
\centerline{\epsfbox{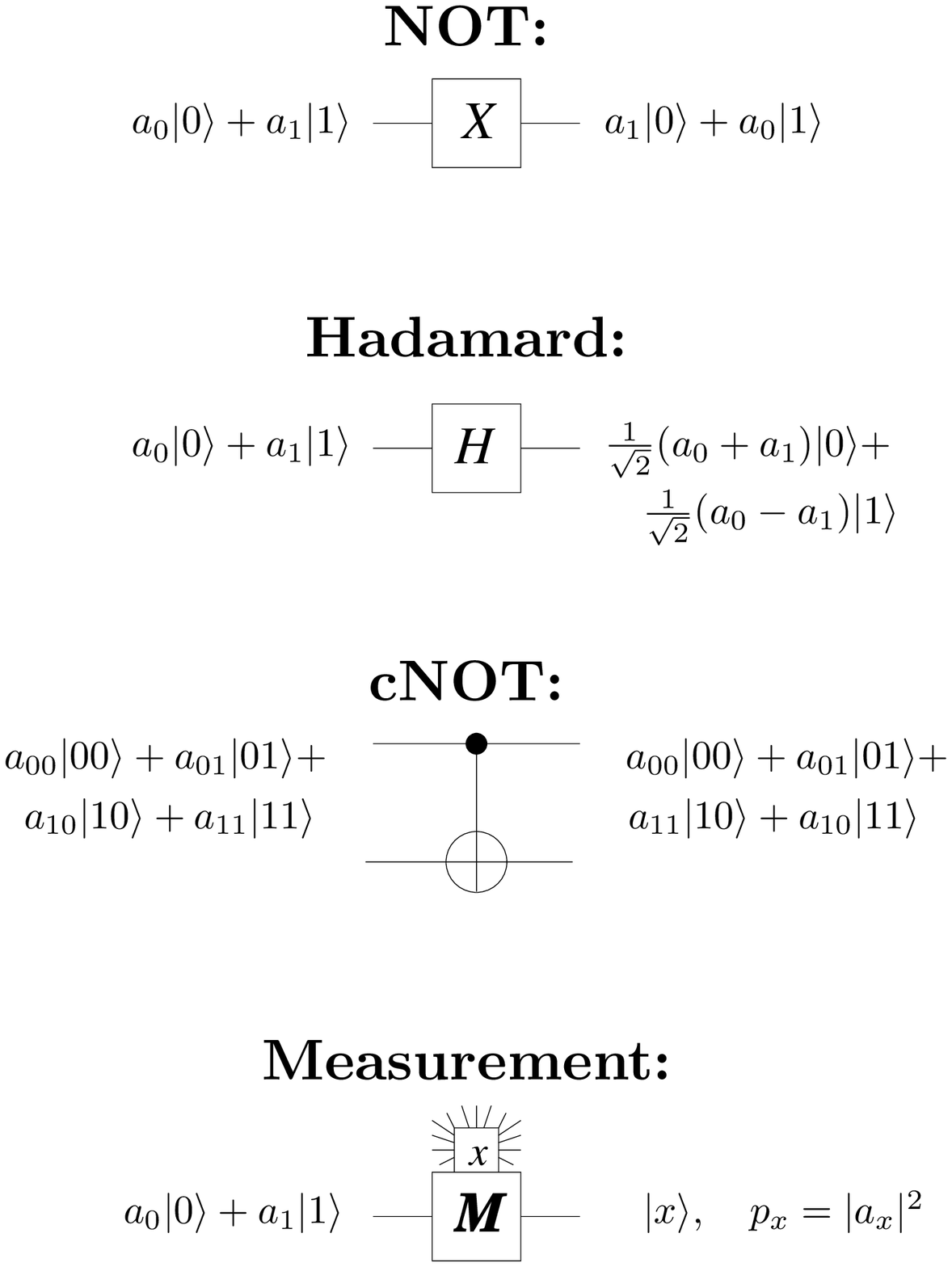}}
\vskip-100pt
\centerline{\bf Figure 1.}

\endinsert

Most importantly, unlike the unitary gates, the measurement gate gives
the user {\it information.}  Although ``information'' is another of
Bell's bad words, there is nothing vague about its use here: the
number $x$ showing in the user-readable display is either 0 or 1.
``Information'' can't get more precise or elementary.

Next in importance, unlike the other gates, a measurement gate has an
ouput characterized by a state that is only statistically determined
by the state associated with its input, the probability $p(x)$ being
given by the modulus squared of the amplitude $a_x$ characterizing the
input state.  

The measurement gate also differs from the others in having no inverse
(a benign example of Bell's bad word {\it irreversibility}).  And its
conversion of incoming to outgoing states is not unitary, and, indeed,
not even linear.

On the other hand the measurement gate {\it resembles\/} the unitary
gates in that the rule governing the state characterizing its output
is completely and precisely specified by the state characterizing its
input.  Another similarity is that the action of the measurement gate,
like that of the unitary gates, is discrete, in contrast to the
notorious distinction traditionally made between the continuous
evolution of the state under the Schr\"odinger equation and the
discontinuous change of state in a measurement.  In most forms of
quantum computation the role of the Schr\"odinger equation is
performed by a small number of discrete unitary gates.  So time
evolution and measurement are alike in both producing discrete changes
of state.

The virtues of the 1-Qbit measurement gate are many, whence my title.
Most obviously, without measurement gates, a computation has no
output.  There is {\it in principle\/} no way, other than reading the
display on measurement gates, for the user of a quantum computer to
extract information from Qbits.  This is less distressing in computer
science than it may be in physics, since there is also no way {\it in
practice\/} for the ordinary user of a classical computer to extract
information from the Cbits, hardwired on microchips buried in the
innards of his or her machine, other than by looking at a visual
display or a print-out.

A less obvious --- or at least less often emphasized --- virtue of the
1-Qbit measurement gate is that, at least in the most straightforward
formulations of quantum computer science, without measurement gates
the computation also has no useful {\it input}, as I shall expand upon
below.  A not unrelated virtue, of great importance for the practical
feasibility of quantum computation, is that without measurement gates
there can be no error correction.

A final virtue, relevant to John Bell's critique of measurement, is
that the many complex, confusing, imprecisely described processes that
go under the name of measurement, requiring the discretion of the
theoretical physicist for their proper interpretation, can all be
unambiguously reduced to elementary 1-Qbit measurement gates, acting in
conjunction with unitary gates, though often not without some
quantum-computational programming ingenuity.  I give a few examples of
this below.

\midinsert
\vskip -30pt
\epsfxsize=6truein
\centerline{\epsfbox{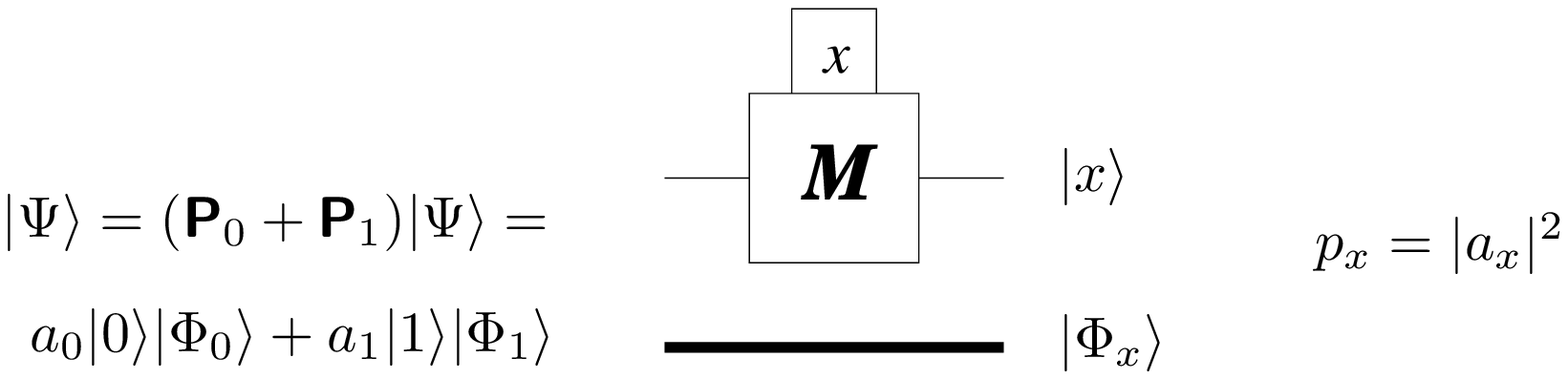}}
\vskip-340pt
\centerline{\bf Figure 2.}
\endinsert

The action of the 1-Qbit measurement gate shown in Figure 1 is
actually a special case of its more general action, when applied to
one of $N$ Qbits, initially in some $N$-Qbit state $\k\Psi$.  Figure 2
gives this more general definition of the 1-Qbit measurement gate.  If
$\Pc_0$ and $\Pc_1$ project a general $N$-Qbit state into its
components along the states $\k0$ and $\k1$ in the subspace associated
with the Qbit on which the measurement gate acts, then the general
$N$-Qbit input state $\k\Psi$ can be expanded as $$\k\Psi =
\Pc_0\k\Psi + \Pc_1\k\Psi = a_0\k0\k{\Phi_0} +
a_1\k1\k{\Phi_1},\eq(NQbit)$$ where $\k{\Phi_0}$ and $\k{\Phi_1}$ are
normalized (but in general nonorthogonal) states of the $N-1$ other
Qbits and $|a_0|^2 + |a_1|^2 = 1.$ If the measurement gate flashes
$x$, then the $N$-Qbit state after the measurement is $\k x\k{\Phi_x}$
with probability $p_x = |a_x|^2.$

That this general definition of the 1-Qbit gate (sometimes called the
``generalized Born rule'') contains as a special case the definition
in Figure 1 (an example of the [ordinary] Born rule), is demonstrated
in Figure 3.  If the single Qbit subject to the measurement gate is
unentangled with the other $N-1$ Qbits, then $\k{\Phi_0} = \k{\Phi_1}
= \k{\Phi}.$ Consequently Figure 2 reduces to Figure 3.  The outgoing
state of the $N-1$ unmeasured Qbits continues to be its incoming state
$\k\Phi$, independent of the reading of the measurement gate, so the
unmeasured Qbits remain unentangled with the measured Qbit.  The state
of the unmeasured Qbits is entirely unaffected by the measurement
gate, while the final state of the measured Qbit is exactly as
specified in Figure 1.

\midinsert
\epsfxsize=6truein
\centerline{\epsfbox{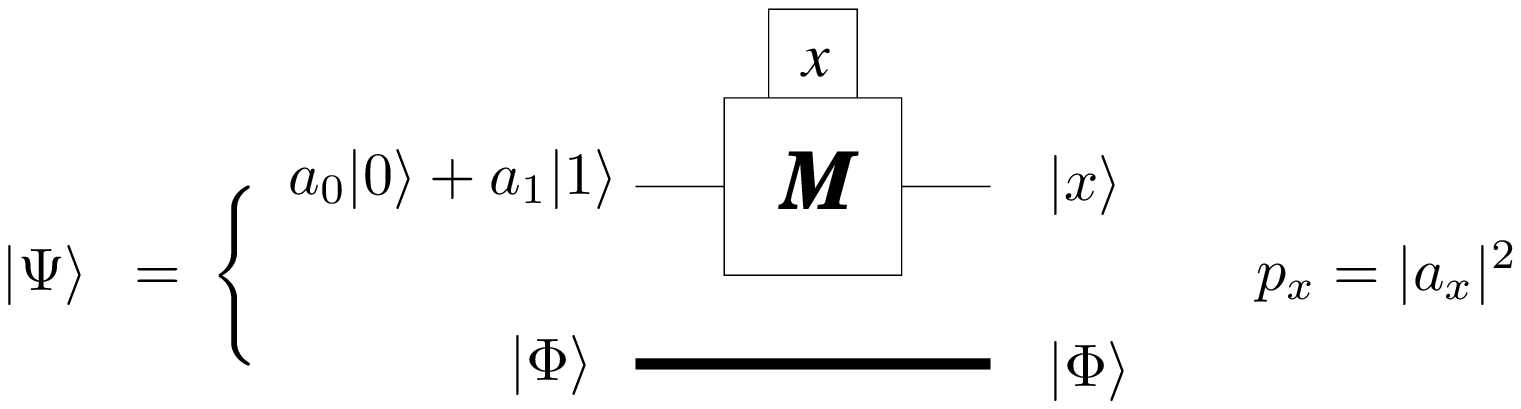}}
\vskip-340pt
\centerline{\bf Figure 3.}
\endinsert

You might think there was need for another kind of measurement
hardware --- a still more general measurement gate that acts on all
$N$ Qbits in the manner specified in the upper half of Figure 4.  But
in fact it can be shown from the generalized Born rule specified in
Figure 3 that precisely this effect is achieved by applying $N$
different 1-Qbit measurement gates to the $N$ individual Qbits,
illustrated for $N=3$ in the lower half of Figure 4.  The proof that
this result always obtains, independent of the order in which the
1-Qbit gates are applied, can be extracted from the generalized Born
rule in a straightforward (but irritatingly clumsy) way.

Finally, all these measurements are what one would call,
more generally, ``measurement in the computational (classical)
basis''.  A more general (von Neumann) measurement on $N$ Qbits would
associate states $\k{\Phi_x}$ with the out-going Qbits taken from some
general complete set of orthonormal states.  Such states can, however,
be unitarily related to the computational-basis states: $\k{\Phi_x} =
\Uc\k x$. Consequently the more general $N$-Qbit measurement gate
$\Mc_{\Uc}$ can be constructed out of the $N$-Qbit computational-basis
measurement gate $\Mc$ as pictured in Figure 5.

\midinsert
\epsfxsize=6truein
\centerline{\epsfbox{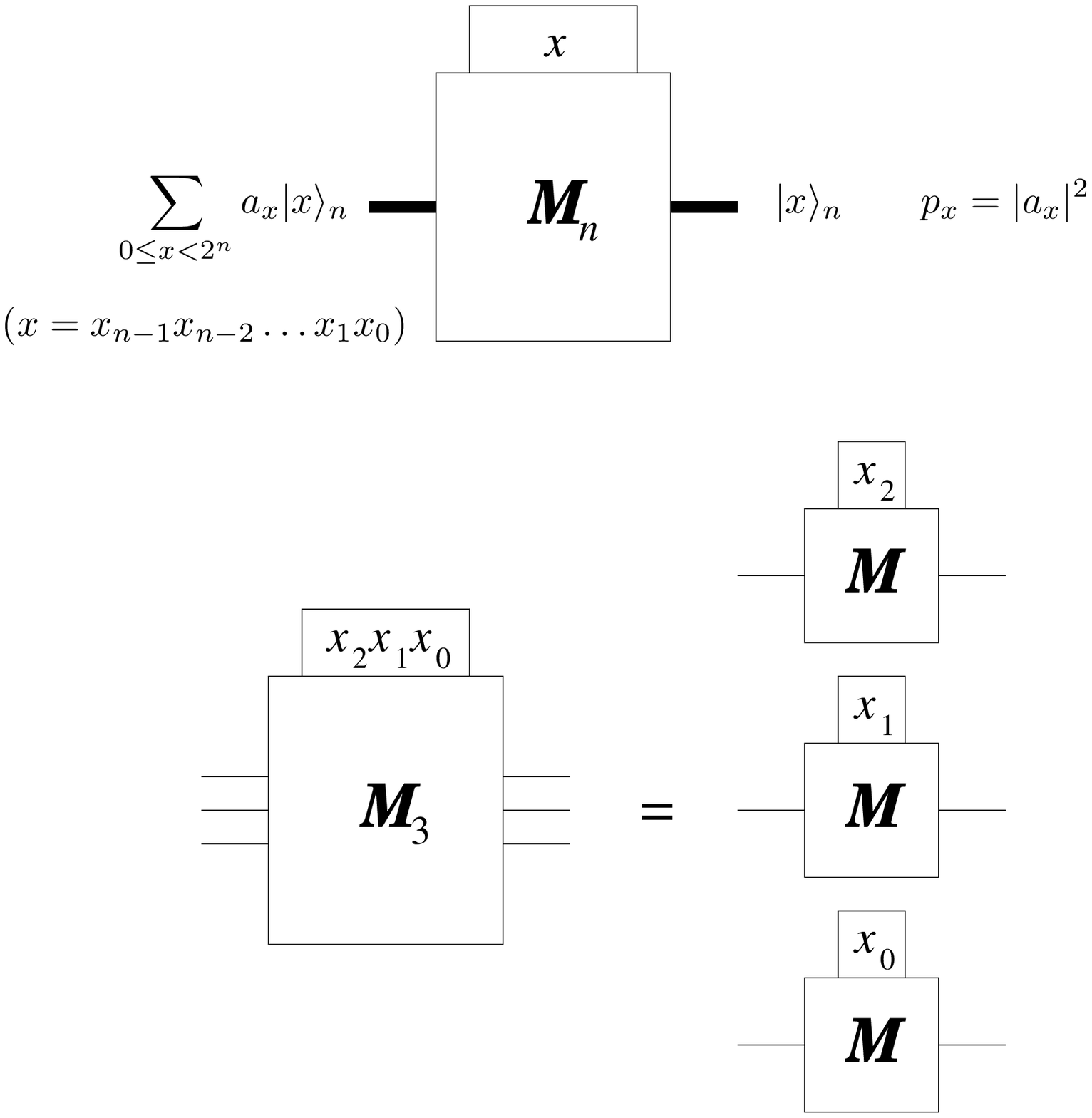}}
\vskip-150pt
\centerline{\bf Figure 4.}
\endinsert

\midinsert

\vskip -20pt
\epsfxsize=6truein
\centerline{\epsfbox{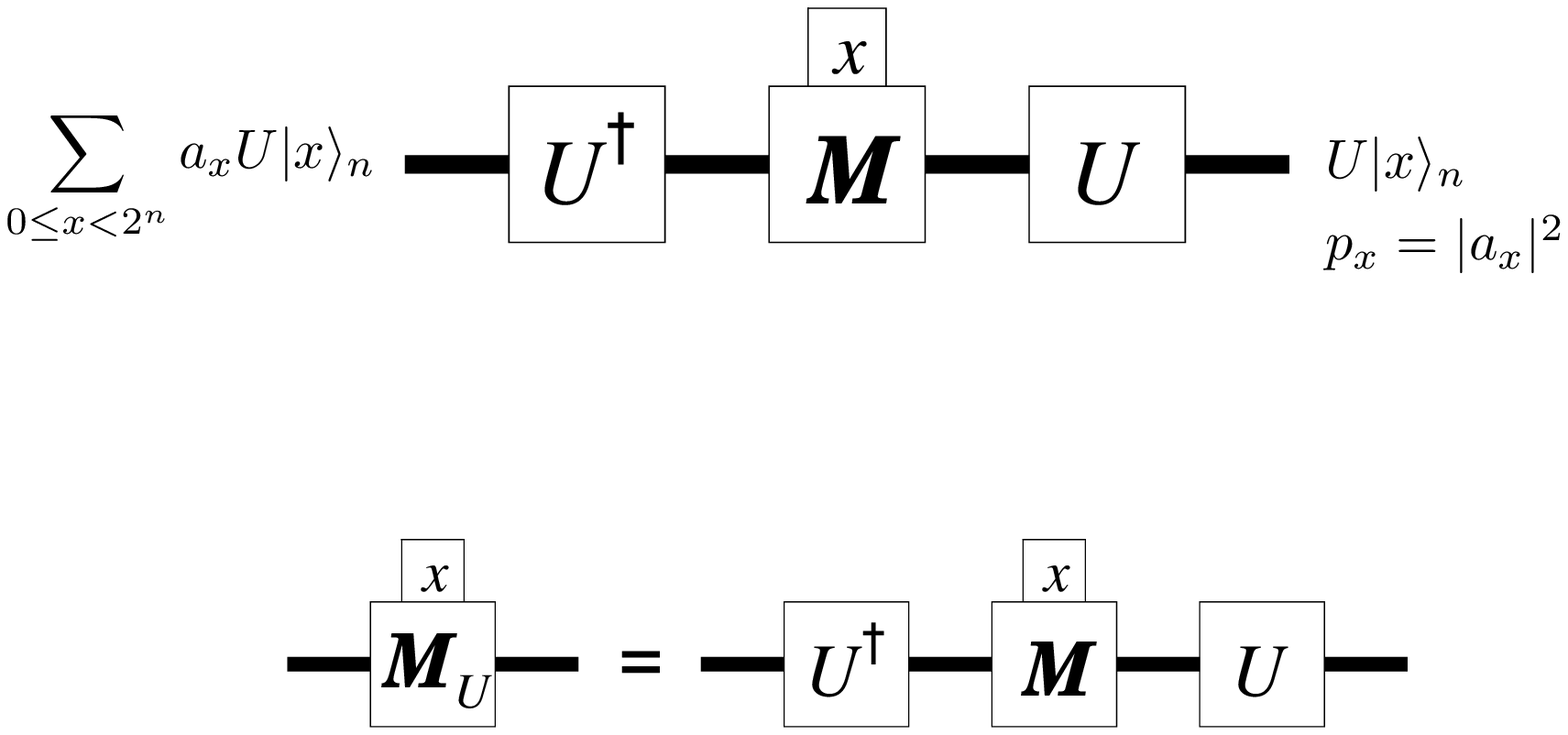}}
\vskip -270pt
\centerline{\bf Figure 5.}
\endinsert

\midinsert

\vskip -30pt
\epsfxsize=7truein
\centerline{\epsfbox{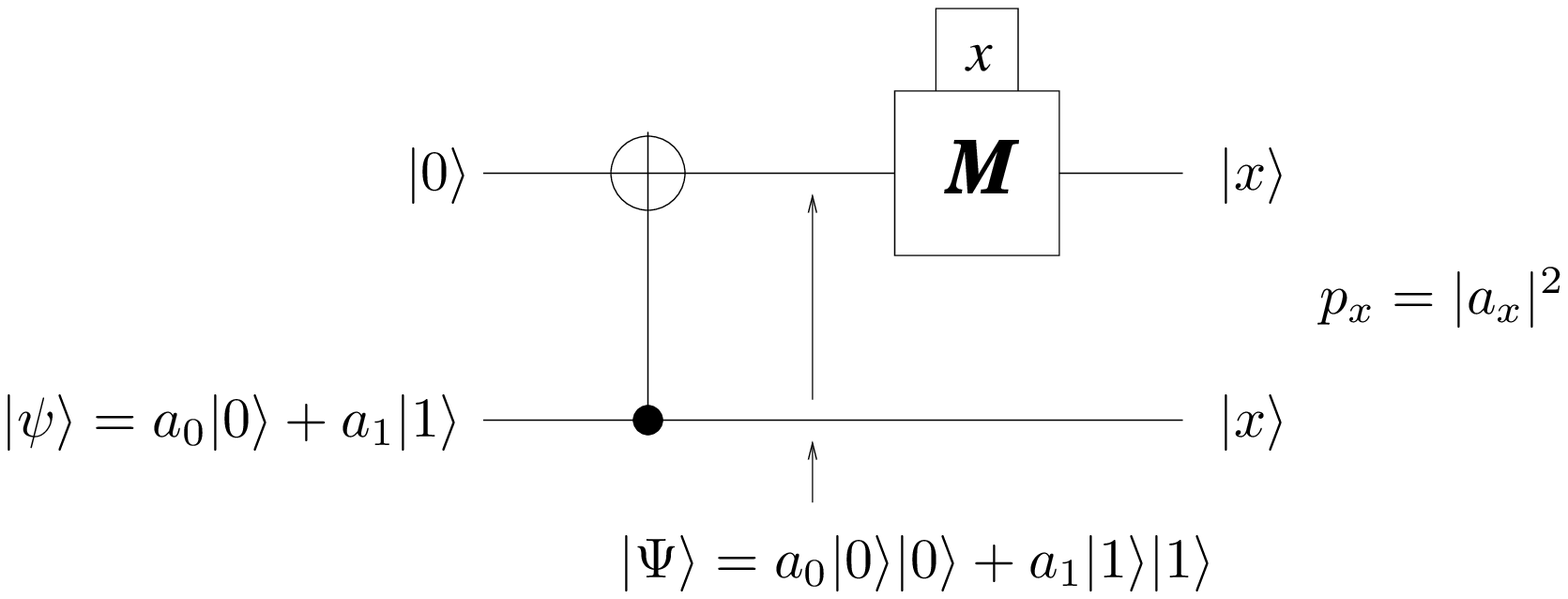}}
\vskip-350pt
\centerline{\bf Figure 6.}
\endinsert

So all von Neumann measurements can be constructed out of unitary
gates combined with one or more specimens of {\it a single black-boxed
piece of hardware\/}: the 1-Qbit measurement gate.  Such obscurity as
there might be in the measurement process rests entirely in that
single elementary gate, whose action is unambiguously specified in
Figure 1.  If there is a ``measurement problem'' in quantum computation,
it is a ``1-Qbit measurement gate problem''.

Note that one can construct an elementary measurement {\it
apparatus\/} --- another on Bell's list of bad words --- out of a
1-Qbit measurement gate, an ancillary Qbit initially in the state
$\k0$, and a cNOT gate that couples the Qbit to be measured to the
ancilla, as pictured in Figure 6.  The measurement gate no longer acts
on the measured Qbit, but only on the ancilla.  Nevertheless, the
resulting final state for the measured Qbit and the associated
probabilities are exactly the same as they would be if the measurement
gate acted on it directly, thanks to its coupling to the ``apparatus''
via the cNOT gate.

$\hphantom{.}$
\topinsert
\epsfxsize=7truein
\vskip -50pt
\centerline{\epsfbox{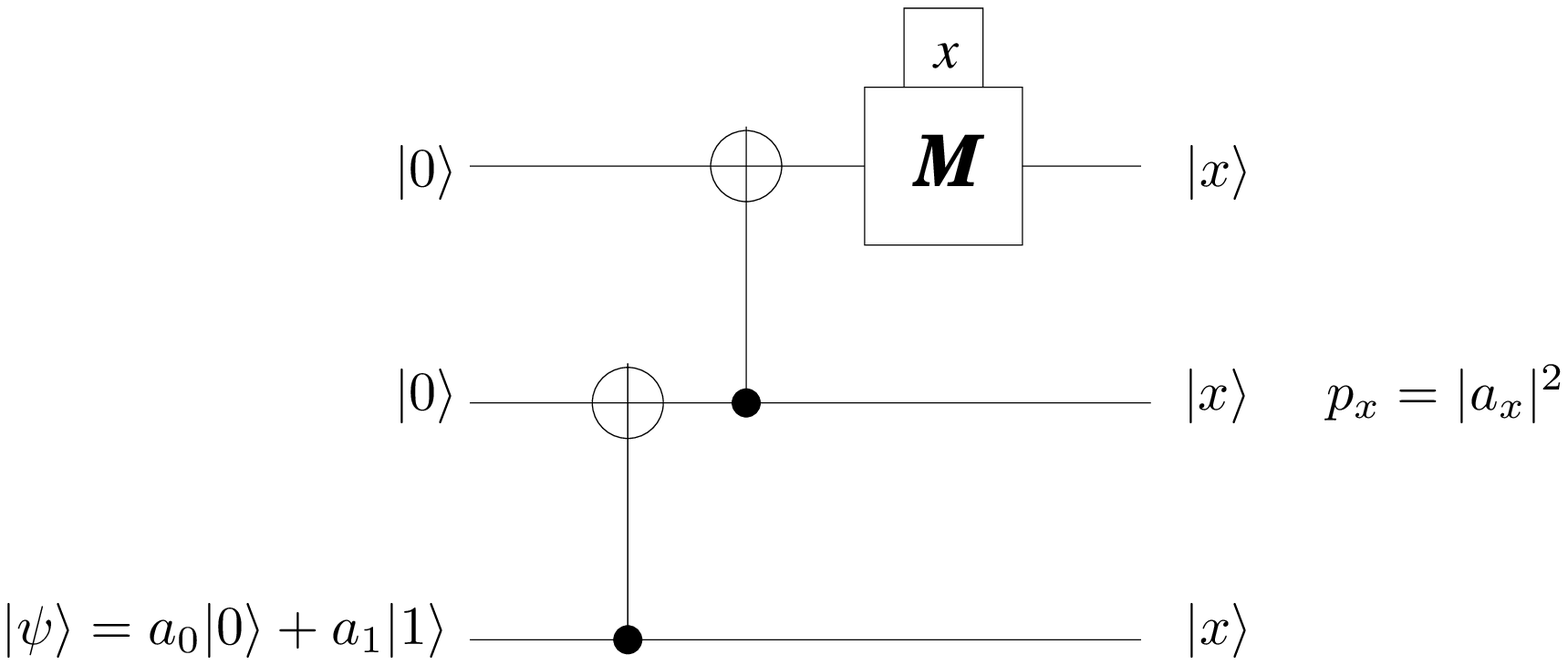}}
\vskip-340pt
\centerline{\bf Figure 7.}
\endinsert

\midinsert
\epsfxsize=7truein
\centerline{\epsfbox{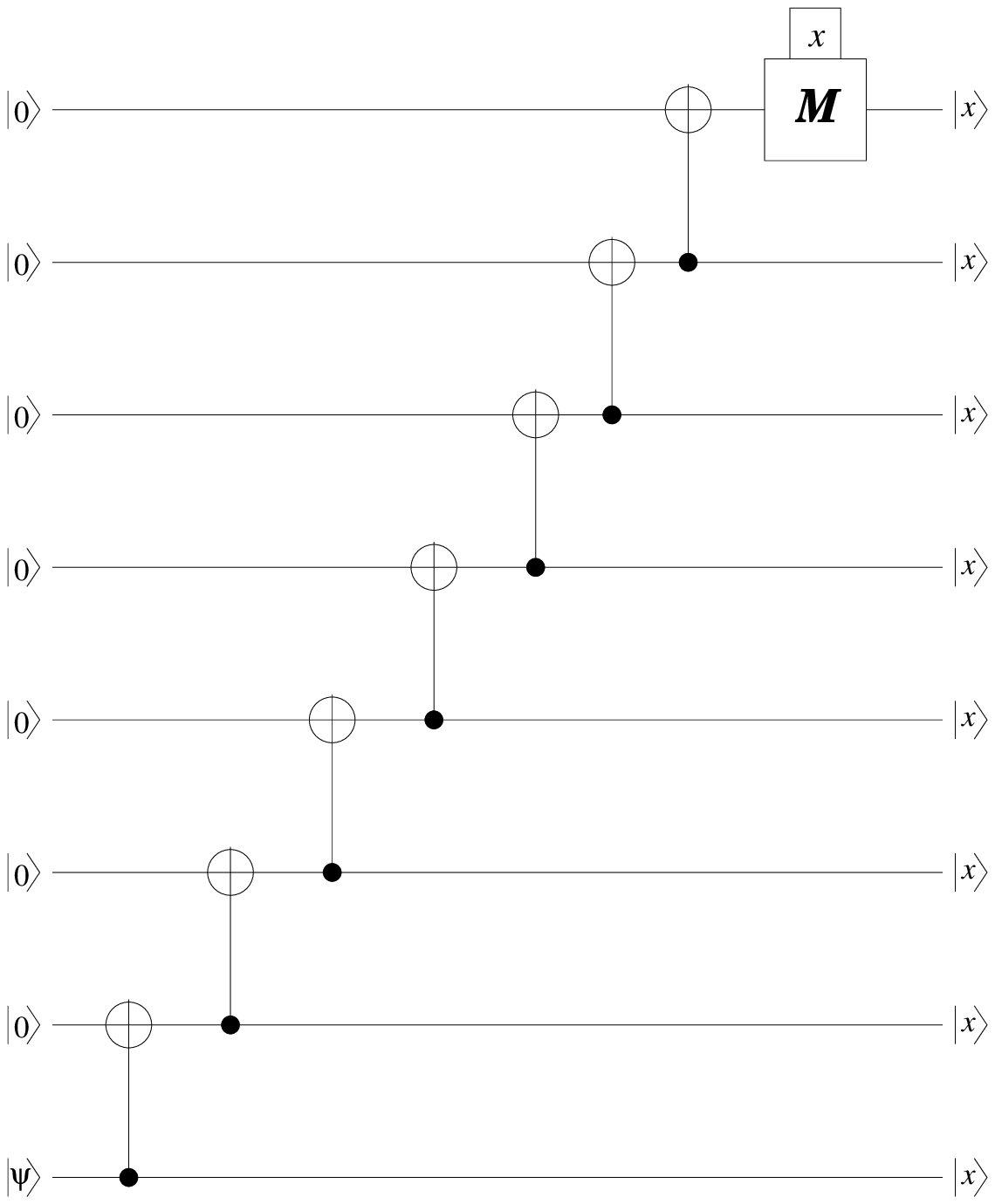}}
\vskip -220pt
\centerline{\bf Figure 8.}
\endinsert

One can also remove the action of the measurement gate from the ancilla, by
introducing a second ancilla, coupled to the first by a second cNOT
gate, as pictured in Figure 7, and can continue in this way for many
stages, as pictured in Figure 8.  At some point the process has to
stop with a measurement gate, if one is to extract any information
from the computer, but one is perfectly free to introduce as many
ancillary Qbits as one wishes, before reaching this stage.  This
corresponds to Bell's ``shifty split'' between what one chooses to
call ``the quantum system'' and what one chooses to call ``the
measurement apparatus'', but there is no longer anything shifty about
it.  A measurement takes place if and only if there is a 1-Qbit
measurement gate somewhere in the circuit.  In the absence of a
measurement gate, since cNOT is its own inverse, one can undo the
whole process and get back to the initial state, as shown in Figure 9.


\midinsert
\vskip 10pt
\epsfxsize=4truein
\centerline{\epsfbox{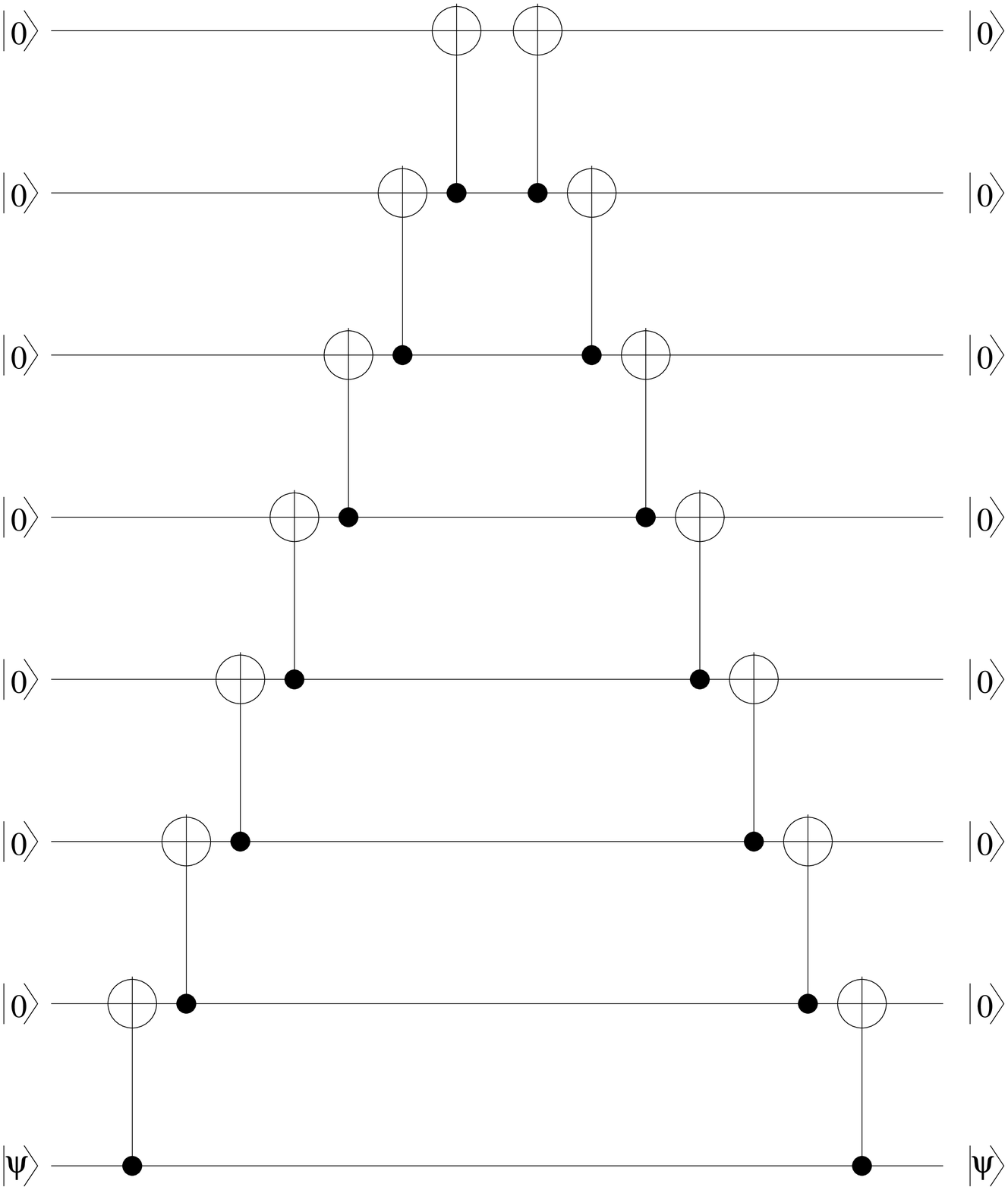}}
\centerline{\bf Figure 9.}
\endinsert


So it is through the readings of 1-Qbit measurement gates, and only
through such readings, that one can extract information about a
computation from a quantum computer.  But this is not the only role
measurement gates play in quantum computer science.  Measurement gates
have a second, equally crucial task to perform, which is
insufficiently emphasized in most expositions of the subject.

The action of all gates, whether they are unitary gates or measurement
gates, is given by a rule that specifies the state associated with the
output Qbits in terms of the state associated with the input Qbits.
But how is a state to be associated with the Qbits at the very
beginning of the computation, before any gates have acted?
The simplest answer sufficient to make a coherent whole of the
subject, and, I would maintain, the only answer that is fully
formulated in mathematical terms, with nothing left to the discretion
of the theoretical physicist, is this: 

Take a Qbit off the shelf, and send it through a 1-Qbit measurement
gate.  If the display on the gate indicates 0, associate the state
$\k0$ with the Qbit emerging from the gate.  If the display indicates
1, associate the state $\k1$.  Thus one can associate the state $\k0$ with
a Qbit by sending it through a 1-Qbit measurement gate and then sending it
through a NOT gate $\Xc$ if and only if the display on the measurement
gate indicates 1.  This is pictured in Figure 10.

\midinsert
\epsfxsize=7truein
\centerline{\epsfbox{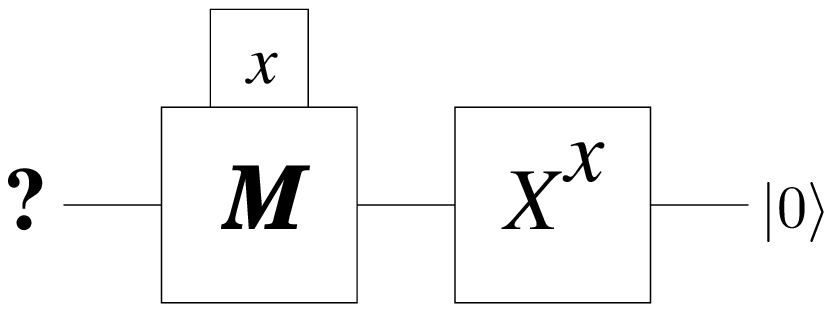}}
\vskip -420pt
\centerline{\bf Figure 10.}
\endinsert

Should the initial Qbit already have a state associated with it, or
should it be a member of a larger group of Qbits which already have a
state associated with them --- i.e. should it be entangled with the
other Qbits in the group --- then Figure 10 is a consequence of the
generalized Born rule for 1-Qbit measurement gates pictured in Figure
2.  If, on the other hand, one has no basis for associating any
initial state either with the Qbit or with a larger group of Qbits to
which it belongs --- even a stochastic association weighted with
probabilities (i.e. a mixed state) --- then the rule illustrated in
Figure 10 must be regarded as an additional property of 1-Qbit
measurement gates, consistent with, but independent of the behavior
specified in Figure 2:

The 1-Qbit measurement gate can thus be used to {\it define\/} what it
means for a Qbit, about whose past history nothing whatever is known, to
be in the state $\k0$.
One can view this role of the 1-Qbit measurement gate in
``initializing a Qbit to the state $\k0$'' as a special kind of cNOT
gate, in which the single Qbit functions as both control (the upper
Qbit in Figure 1) and target (the lower Qbit in Figure 1).  Such an
auto-erotic cNOT gate, unlike a normal 2-Qbit cNOT gate, is not
unitary, since the state associated with its output is $\k0$,
independent of whether the state associated with its input is $\k0$ or
$\k1$.

\midinsert
\epsfxsize=5truein
\vskip -30pt
\centerline{\epsfbox{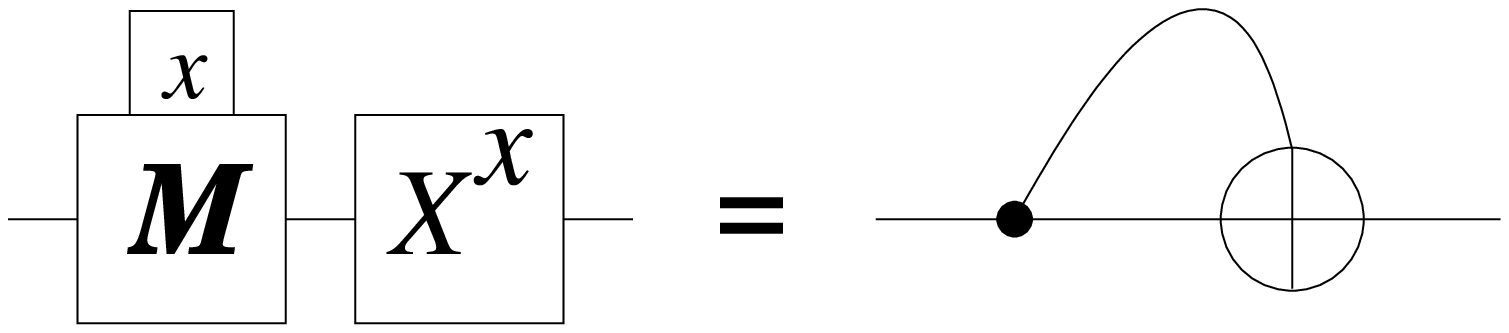}}
\vskip -300pt
\centerline{\bf Figure 11.}
\endinsert

\bigskip

Up until now I have used the term ``state'' without actually defining
it in mathematical terms (with nothing left to the discretion of the
theoretical physicist).   But now we can give a precise definition:

The state $\k x$ ($x=0$ or 1) is associated with a Qbit if the Qbit,
perhaps in association with other Qbits, has been subjected to a
series of gates whose actions, together with the readings of any gates
that are measurement gates, imply, by the rules of gate operation
specified above, that the state associated with the Qbit is $\k x$. 

A particularly simple example of those rules is that the state $\k x$
is associated with a single Qbit emerging from a 1-Qbit measurement
gate that reads $x$.
  
More generally, to associate the state $\k\Psi$ with $N$ Qbits means
nothing more nor less than that the Qbits, perhaps in association with
other Qbits, have been subjected to a series of gates whose actions,
together with the readings of any measurement gates, imply that the
state associated with the Qbits is $\k\Psi.$

The reader may have noted my practice of replacing the customary
phrase ``state of the Qbits'' with ``state associated with the
Qbits''.  The reason for this awkward expansion is that the simpler
phrase, to the extent that it suggests that the state resides in or is
an inherent property of the Qbits, is incorrect.  Given full access
to the Qbits, but no information about them, there is, famously,
nothing you can do to them to inform you of their associated state.
That state is nothing more than a compact summary of all features of
their past history relevant to determining the probabilities of the
readings of subsequent measurement gates, applied either immediately
or after the action of additional unitary gates.

Unlike the misleading phrase ``state of the Qbits'', the phrase
``state associated with the Qbits'' immediately raises the question
``associated by whom?''  Quantum computer science provides a
straightforward answer to this question: the user.  Here ``user'' is
the standard computer-science user, most commonly encountered in the term
``user-friendly''.  It should be clear that ``user'' does not mean God
(who already knows the factors of all integers and therefore does not
need to use a quantum computer to help him out).  Nor does it mean the
mouse that Einstein worried about, since no mouse (even Einstein's)
has any interest in factoring large integers.  ``The user'' is nothing
more or less than the user.

Having made this important point, I will now use the more graceful
phrase ``state of the Qbits'' with the understanding that ``of'' is
simply a compact abbreviation of ``associated with'' or ``assigned
to''.

\midinsert
\vskip -40pt
\epsfxsize=7truein
\centerline{\epsfbox{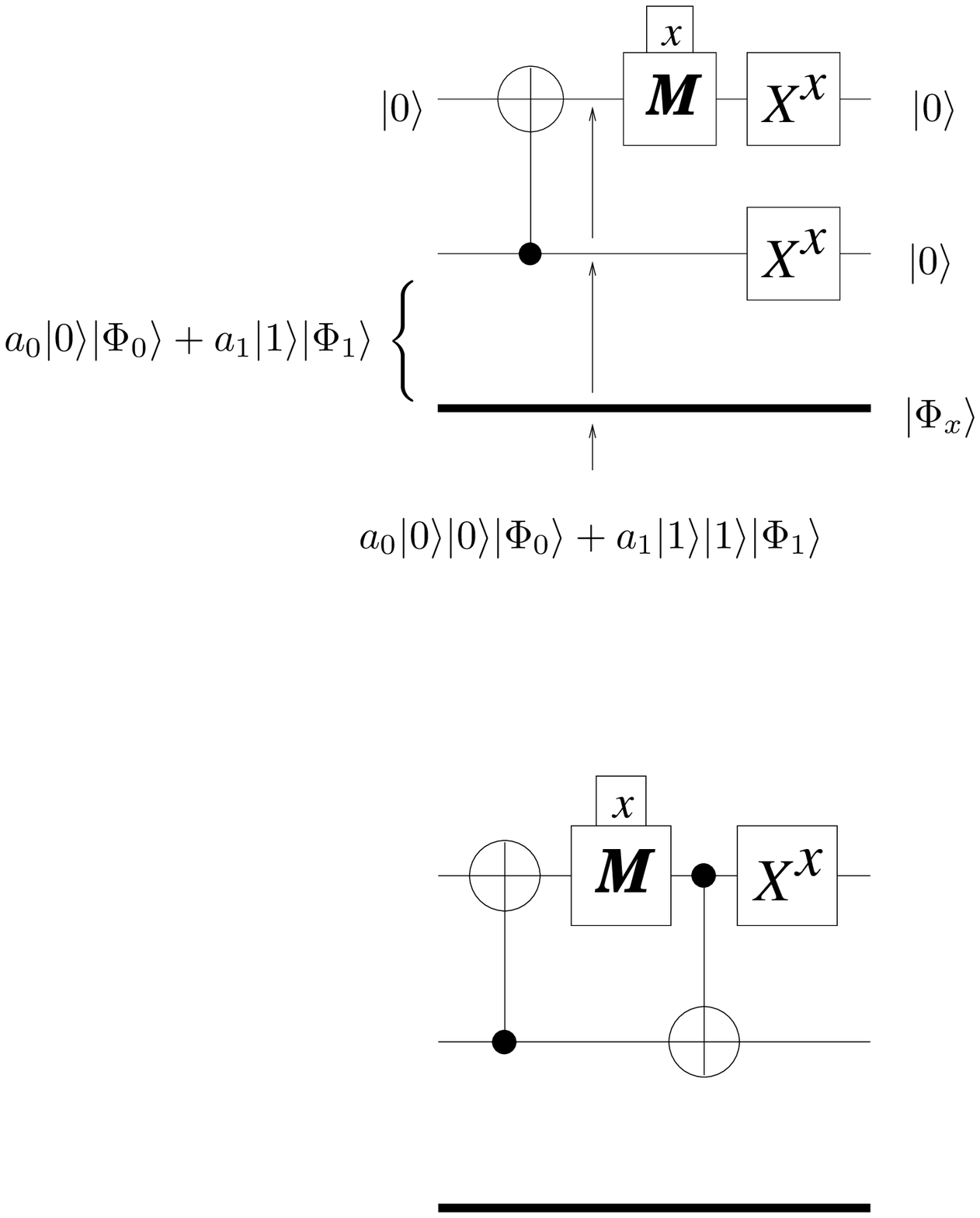}}
\vskip -110pt
\centerline{\bf Figure 12.}
\endinsert


Figures 12 and 13 explore the question of whether it might be possible
to eliminate this use of measurement gates in assigning the state
$\k0$ to an initially stateless Qbit.  The Qbit in question is
associated with the middle wire in Figure 12.  Its initial
statelessness is taken to arise from its entanglement with many other
``external'' Qbits, associated with the heavier bottom wire.  The
entanglement is characterized by a state
$a_0\k0\k{\Phi_0}+a_1\k1\k{\Phi_1}$, where $\k0$ and $\k1$ are
associated with the Qbit and $\k{\Phi_0}$ and $\k{\Phi_1}$ are
associated with the Qbits with which it is entangled.

Rather than applying a measurement gate directly to the Qbit to
associate with it the state $\k0$ as in Figure 10, we introduce an
ancilla (top wire in Figure 12) coupled to the Qbit with a cNOT gate,
as in the measurement apparatus of Figure 6. (Throughout the rest
of this discussion I refer to the Qbit on which the state is to be
conferred as ``the Qbit'' and the ancillary Qbit, as ``the ancilla.'')
By applying or not applying $\Xc$ to the Qbit, depending on whether or
not the measurement gate applied to the ancilla indicates 1 or 0, we
can associate with the Qbit the state $\k0$ even though no measurement
gate has been applied to it directly, though, of course, a measurement
gate (applied to the ancilla) has still played a crucial role in this
state assignment.

We can automate this process, replacing the NOT gate that acts or
doesn't act on the Qbit, depending on the reading of the measurement
gate, by a cNOT controlled by the ancilla and targeted on the Qbit.
This is pictured in the lower part of Figure 12 and the upper part of
Figure 13.

\midinsert
\epsfxsize=7truein
\centerline{\epsfbox{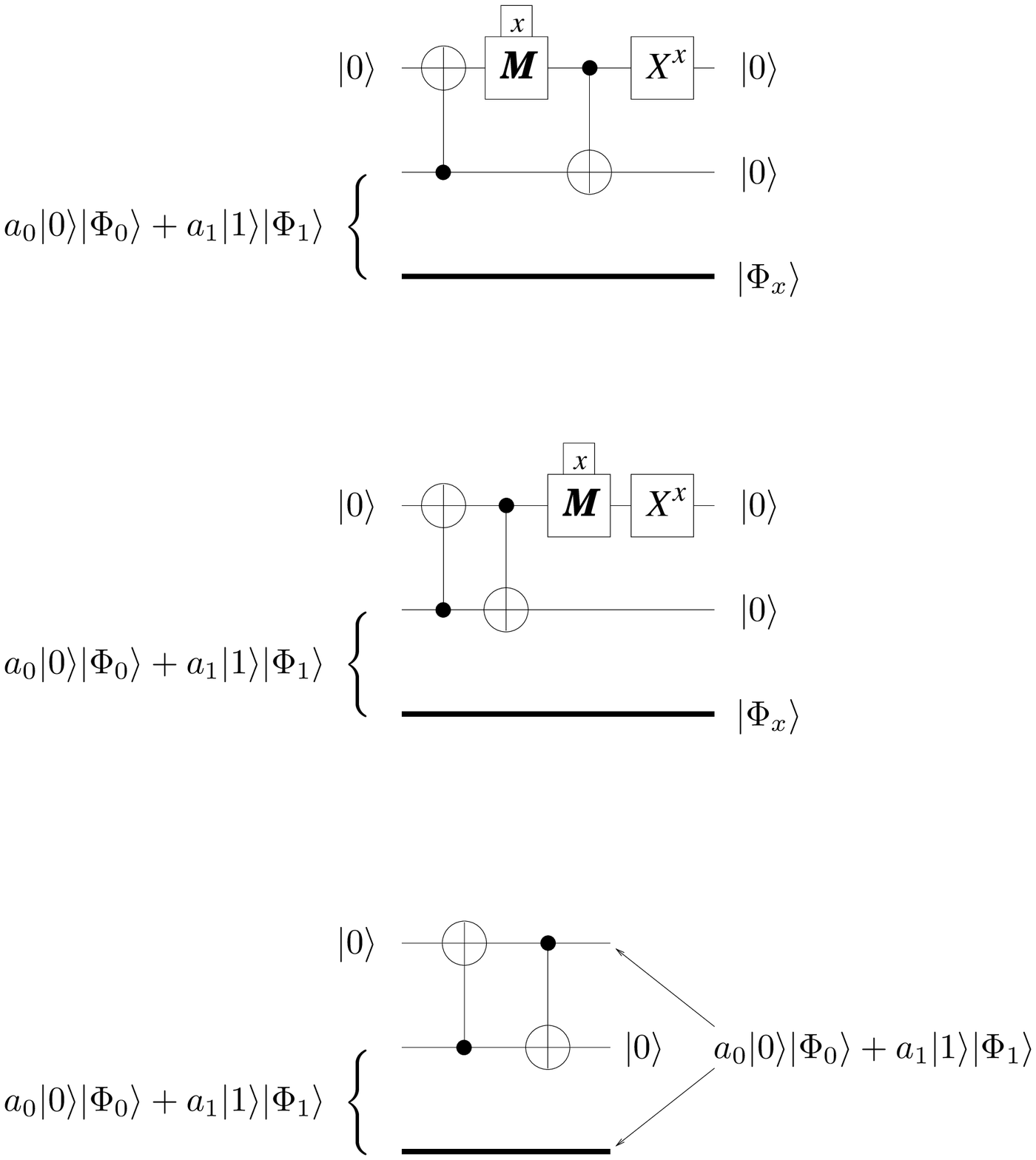}}
\vskip -80pt
\centerline{\bf Figure 13.}
\endinsert

But now we can use the (easily verified) fact that the combined
outcome of a cNOT gate followed by a measurement gate on the target
Qbit is unaltered by changing the order in which the two gates act.
This gives us the middle part of Figure 13, in which the measurement
gate only acts {\it after\/} the Qbit has been assigned the state
$\k0$.  The measurement gate (and the NOT gate that follows it if the
reading of the measurement gate is 1) can therefore be dropped,
without altering the state assignment to the Qbit, leaving us with the
bottom part of Figure 13.  So we have managed to associate the state
$\k0$ with the Qbit, without having to make any use of a measurement
gate.  The measurement gate has been eliminated from the act of state
preparation.

But to do this we required an ancilla in the state $\k0$.  Indeed, all
the circuit at the bottom does is swap the roles played by the ancilla
and the Qbit.  The Qbit acquires the state $\k0$ initially possessed
by the ancilla, and the ancilla ends up entangled with the external
Qbits in exactly the same way that the Qbit originally was.  If we
wish to restore the ancilla to its initial state, so that we have
actually produced an {\it additional\/} Qbit associated with the state
$\k0$ rather than merely swapped the state $\k0$ from one Qbit to
another, then we must retain the final operations on the ancilla
pictured in the middle part of Figure 13.


The situation here is reminiscent of Charles Bennett's solution [3] to
the problem posed by Maxwell's demon.  In the bottom part of Figure 13
we do indeed manage to get the Qbit into the pure state
$\k0$, but at the price of removing from the demon (ancilla) its own
pure state.  To get the demon back to a condition in which it can
continue its work, it is necessary to erase the entanglement it has
acquired with the external Qbits (by measuring it), and we have
therefore failed in our attempt to eliminate measurement gates from
the act of state preparation.

It is much the same with error correction, which is, in a sense,
nothing more than a refined form of state preparation.   To illustrate
this, and to elaborate on yet another way in which measurement is
unproblematic (as well as crucial) in quantum computation, it is
necessary to say a little about the measurement of degenerate
observables.  

\midinsert

\vskip -45pt
\epsfxsize=7truein
\centerline{\epsfbox{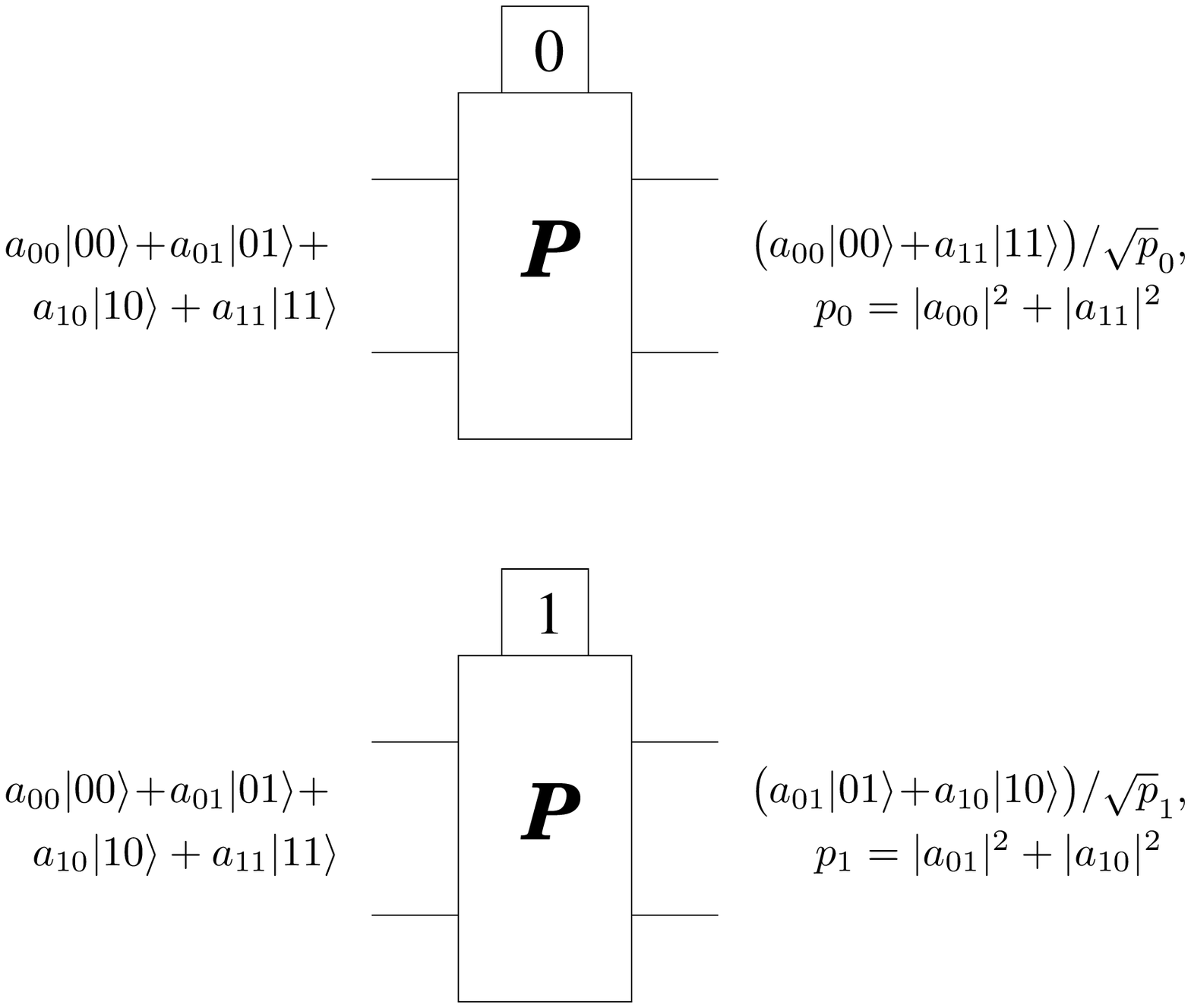}}
\vskip -200pt
\centerline{\bf Figure 14.}
\endinsert

We have already seen an example of measuring a degenerate observable
in the generalized Born rule, illustrated in Figure 2, the observable
being either of the projection operators $\Pc_x$ on the computational
basis states $\k x$ of the measured Qbit, each of which is
$2^{N-1}$-fold degenerate in the $2^N$-dimensional space of $N$ Qbits.
Devising ways to measure less trivial degenerate observables with
1-Qbit measurement gates can require a certain amount of quantum
programming ingenuity, but this is surely what Bell meant when he
deplored leaving things to the discretion of the theoretical
physicist.

Figure 14 shows, as an important example, the degenerate 2-Qbit
parity-measurement gate $\Pc$, which outputs, with the
appropriate probabilities, the normalized projection of the input
2-Qbit state into the even and odd parity subspaces, spanned by
$\k{00}$ and $\k{11}$ (even) or $\k{10}$ and $\k{01}$ (odd).  Figure
15 shows how to realize this gate using a single 1-Qbit measurement
gate and the aid of an ancilla, as usual initialized to the state
$\k0$, which is coupled by cNOT operations to each of the two Qbits
whose parity is being measured.  Since the combined effect of the two
cNOT gates on the ancilla is to leave it in the state $\k0$ if the two
measured Qbits are in either of the even parity computational basis
states, and flip it to the state $\k1$ if they are in either of the
odd states, it follows from linearity and the generalized Born rule
that this circuit achieves the parity measurement of Figure 14, as
summarized in Figure 16.  (The two final cNOT gates act to restore the
ancilla to its initial state.)

\topinsert
\vskip -45pt
\epsfxsize=7truein
\centerline{\epsfbox{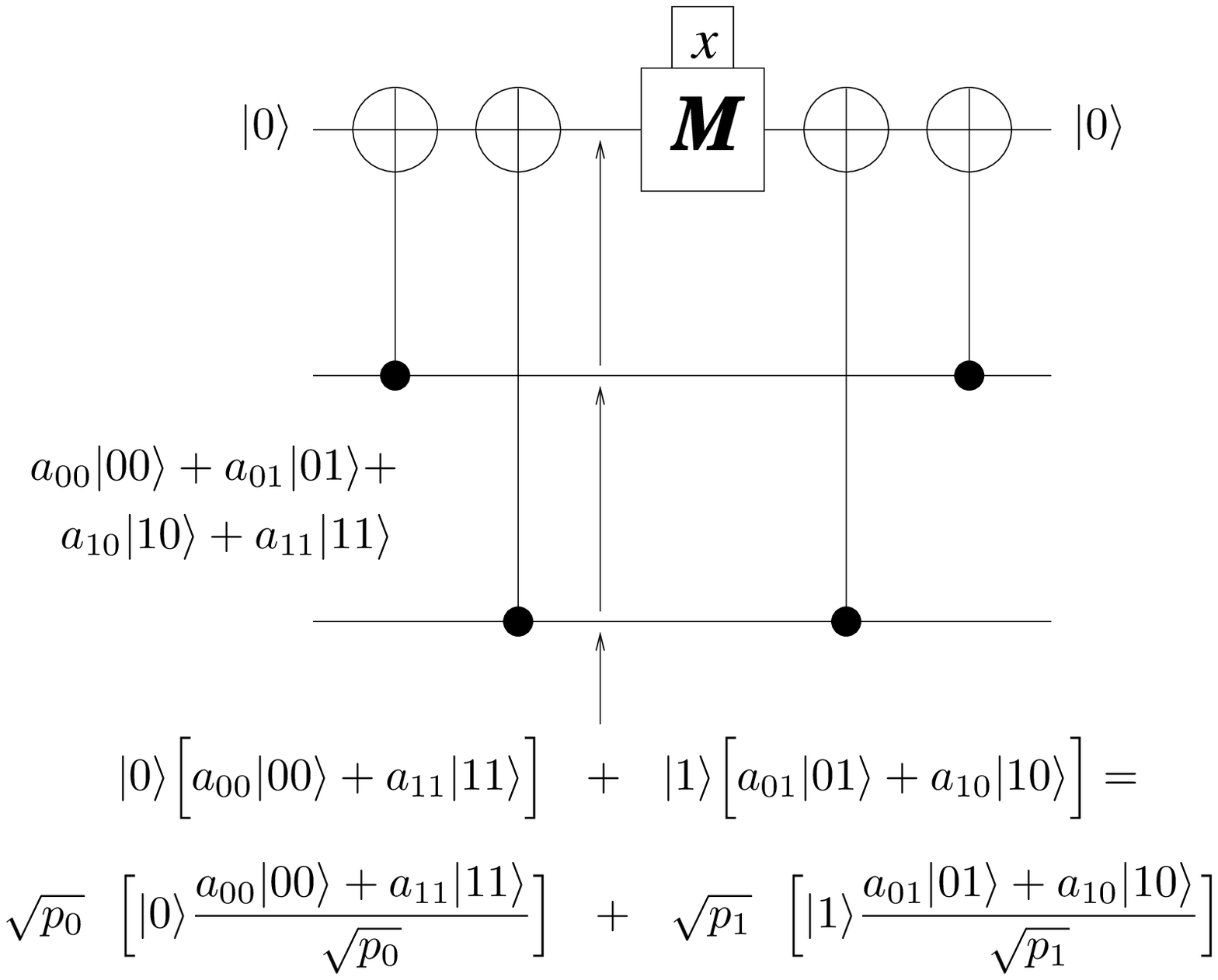}}
\vskip -225pt
\centerline{\bf Figure 15.}

\epsfxsize=7truein
\centerline{\epsfbox{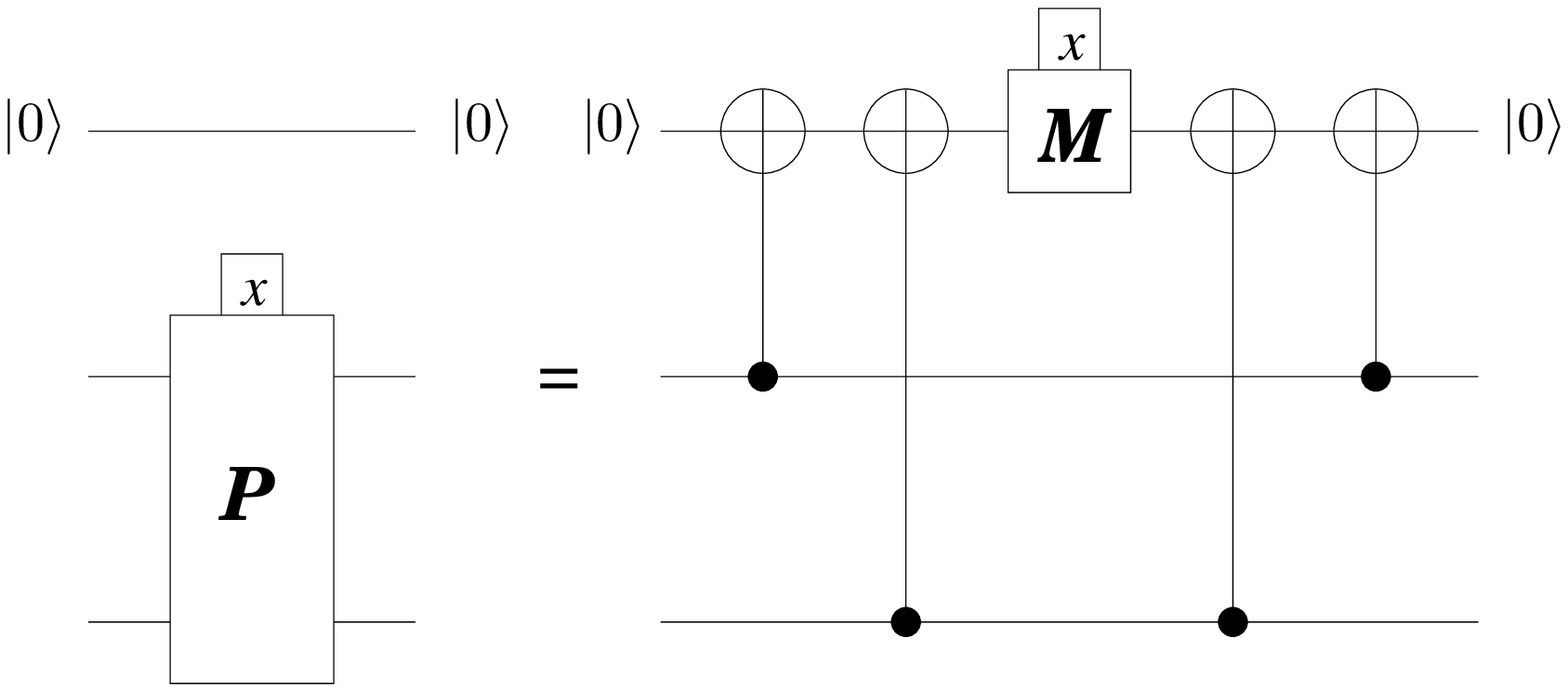}}
\vskip -350pt
\centerline{\bf Figure 16.}
\endinsert

This is one of the simplest examples of measuring a degenerate
observable using ancillas and 1-Qbit measurement gates.  A more subtle
example is shown in Figure 17.  The upper part shows a
circuit that measures the total spin of two spin-$\fr1/2$ Qbits (with
the understanding that the 1-Qbit state $\k0$ represents spin up, and
$\k1$, spin down).  The central 3-Qbit gate is a doubly-controlled NOT
(or Toffoli) gate, which, acting on the computational basis, flips the
state of the target Qbit if and only if both control Qbits are in the
state $\k1$.  In quantum-information-theoretic language, the
post-measurement state associated with the two Qbits is the
anti-symmetric Bell state $\fr1/{\sqrt 2} \lp\k0\k1 -\k1\k0\rp$ (total
spin 0) if the reading $s$ of the measurement gate is 0, and is the
projection of its initial state into the 3-dimensional symmetric
subspace (total spin 1), if the measurement gate indicates 1.

To see that the circuit behaves in this way, note first that if $\Ac$
is any $N$-Qbit unitary gate, and an ancilla acts as a control Qbit for a
controlled-A gate, then if the state of the ancilla is $\Hc\k1$ prior
to the action of the controlled-A gate, where $\Hc$ is the Hadamard
gate of Figure 1, and a second Hadamard gate is applied to the ancilla
after the controlled-A gate acts, then the final state of the $N+1$
Qbits is easily confirmed to be the one shown in the lower part of
Figure 17.

\midinsert
\epsfxsize=7truein
\centerline{\epsfbox{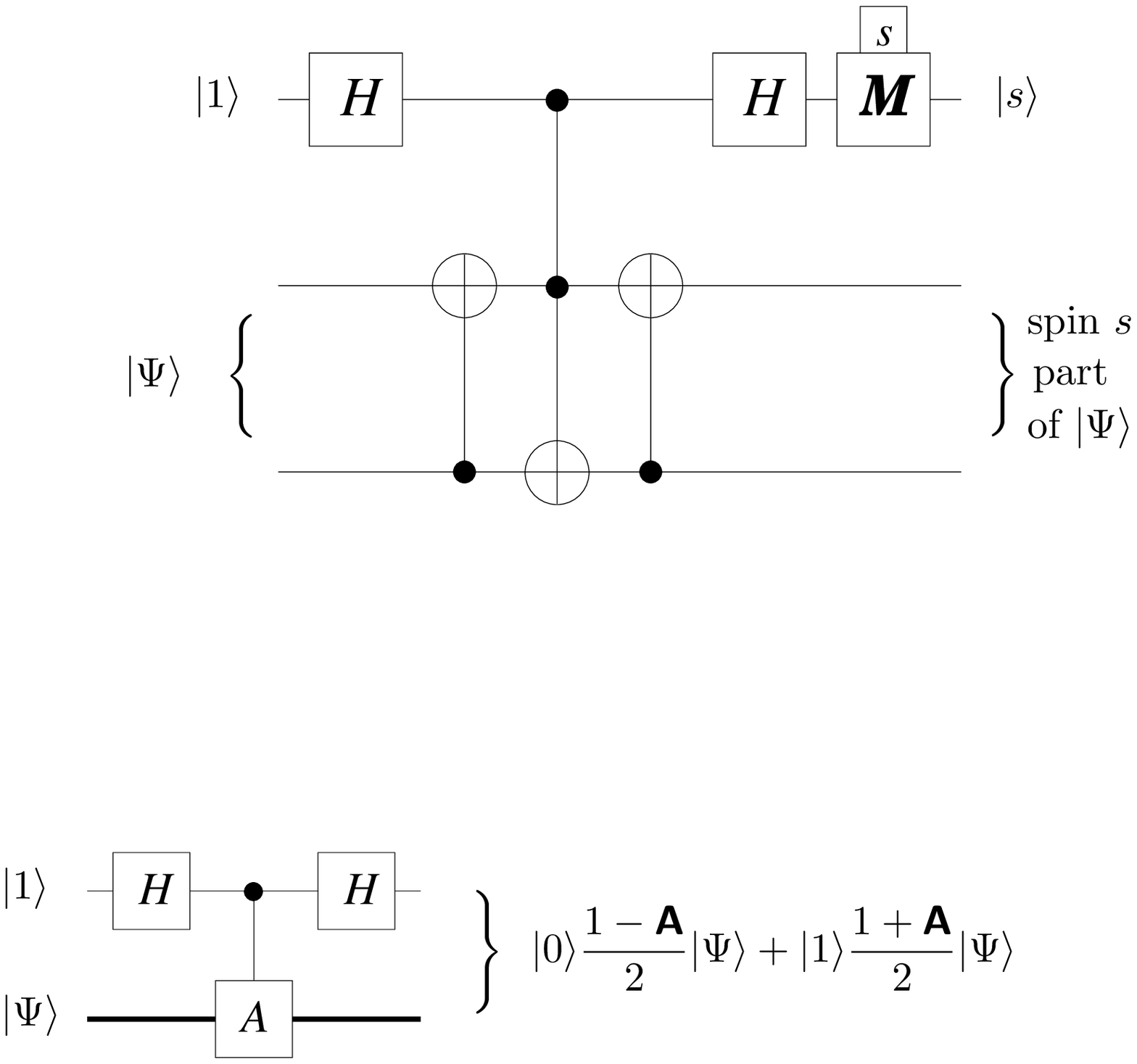}}
\vskip -160pt
\centerline{\bf Figure 17.}
\endinsert

The operation between the two Hadamards in the upper part of Figure 17
is precisely of this form (for $N$=2) with $\Ac$ being a set of three
cNOT gates (with alternating control and target Qbits) acting on the
lower two Qbits.  Such a set of three alternating cNOT gates is
easily confirmed to be the SWAP gate, which, acting on the
computational basis state $\k x\k y$ gives $\k y\k x$.  But when $\Ac$
is a two-Qbit swap gate, then $\h\lp\oc-\Ac\rp$ projects on the
(1-dimensional) antisymmetric subspace, while  $\h\lp\oc+\Ac\rp$
projects on the (3-dimensional) symmetric subspace.  

Therefore the generalized Born rule requires the final state
associated with the two lower Qbits to be the symmetric or 
anti-symmetric component of the input state, depending on whether the
measurement gate acting on the ancilla reads 0 or 1.

An amusing feature of this example (and many other quantum computational
circuits) is that there is quite a different way to understand why it
behaves as it does.  In this case one can construct an alternative
view of the circuit by noting that if any control Qbit in a
multiply-controlled NOT gate is sandwiched between Hadamards, then,
without changing the action of the circuit, the control Qbit can be
changed to a target Qbit (without the two Hadamards) provided the
former target Qbit is changed to a control Qbit sandwiched between two
Hadamards.  (This follows most simply from the fact that $\Xc =
\Hc\Zc\Hc$, where $\Zc$ is the phase gate, $\Zc\k x = (-1)^x\k x$, and
the fact that the action of a multiply-controlled $\Zc$ gate is
unaltered by interchanging the target Qbit with any of the control
Qbits, since all it does is multiply the $(N+1)$-Qbit computational
basis state by $-1$ if and only if that states is a product of $N+1$
states $\k1$.)

\midinsert
\epsfxsize=7truein
\centerline{\epsfbox{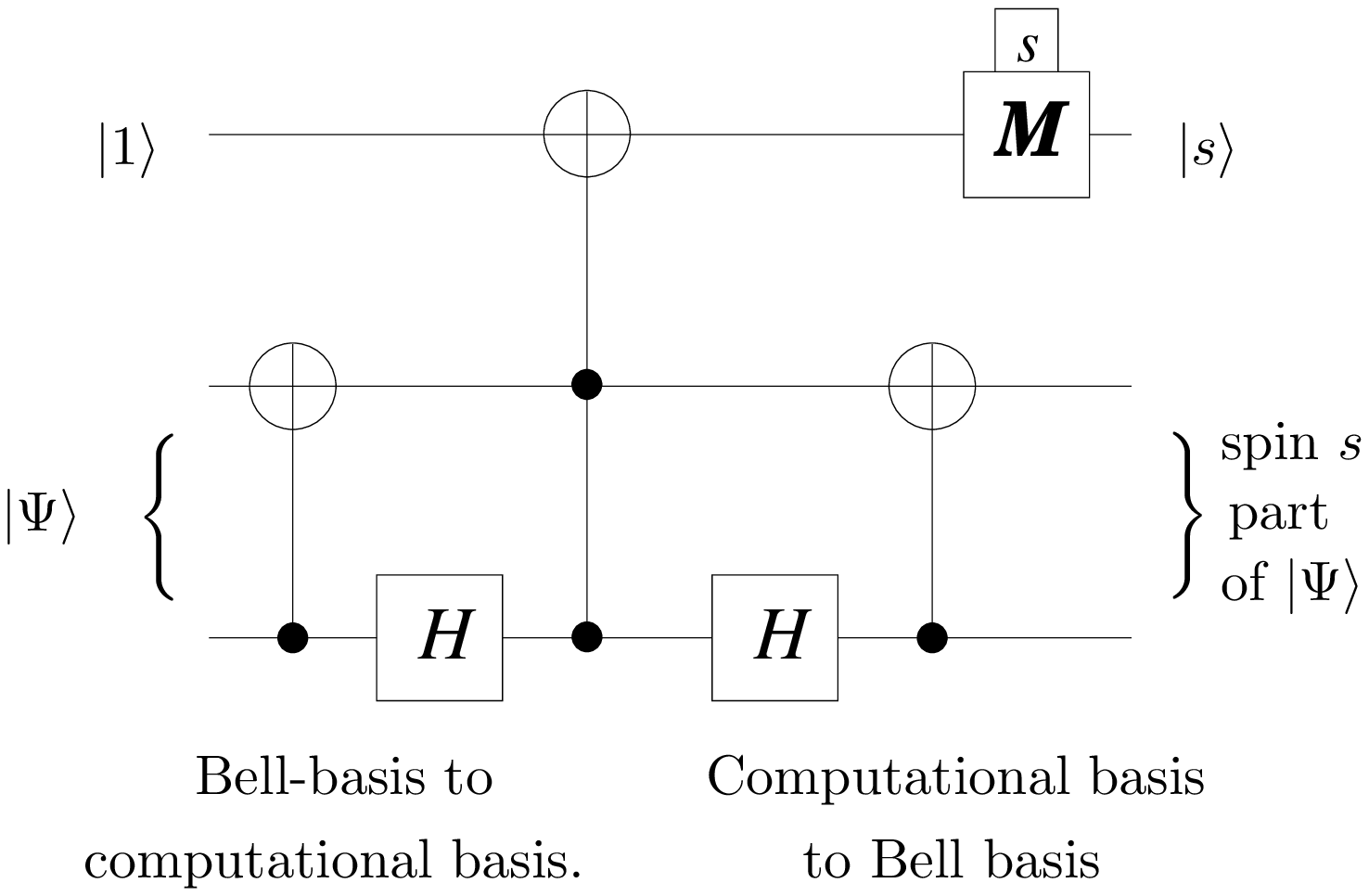}}
\vskip -270pt
\centerline{\bf Figure 18.}
\endinsert

Figure 18 shows the upper part of Figure 17, modified in precisely
this way.  But the action of this circuit has quite a different
interpretation.  The first two gates on the left act on the lower two
Qbits to convert the Bell basis to the computational basis.  In
particular the (singlet) Bell state $\fr1/{\sqrt2}\lp\k0\k1-\k1\k0\rp$
is taken into the computational basis state $\k1\k1$, while the three
symmetric (spin 1) Bell states are taken into the computational basis
states $\k1\k0, \k0\k1$, and $\k0\k0$.   So if $\k\Psi$ is the singlet   
state, then the NOT acts on the ancilla and the measurement gate registers
0, but if $\k\Psi$ is any triplet (spin-1) state, the NOT does not act
and the measurement gate registers 1.

I conclude with a glimpse into the role played by measurement in
error correction, emphasizing the conceptual similarity to state
preparation.  Suppose we encode any 1-Qbit state $\al\k0 + \be\k1$
into a 3-Qbit code state $\al\k0\k0\k0 + \be\k1\k1\k1$ and suppose
(artificially) that the only corruption that can befall the three
Qbits is that at most one of them may be subject to a NOT gate (which
is the general classical single-Cbit error, but not the general
quantum single-Qbit error.)  We require a circuit that tests for such
an error and, if a bit-flip error is found, restores the original
encoded state $\al\k0\k0\k0 + \be\k1\k1\k1$, whatever the values of
the amplitudes $\al$ and $\be$.

\midinsert
\epsfxsize=7truein
\centerline{\epsfbox{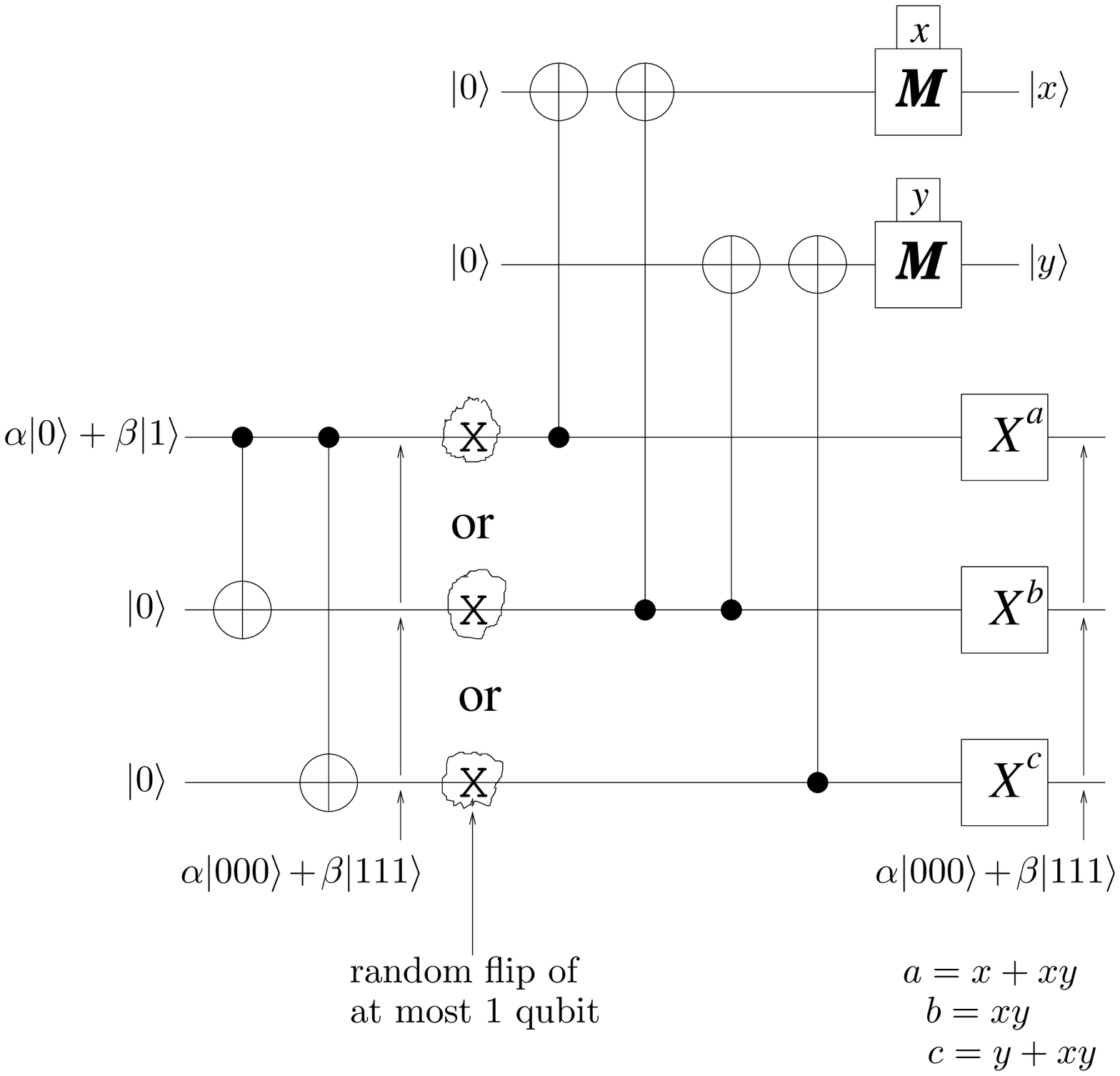}}
\vskip -100pt
\centerline{\bf Figure 19.}
\endinsert

Figure 19 shows the entire process of error production and correction.
On the lower left we have the unencoded 1-Qbit state $\al\k0 + \be\k1$
and two additional Qbits in the state $\k0$.  The two initial cNOT
gates on the lower left act to transform these three Qbits into the
3-Qbit code state $\al\k0\k0\k0 + \be\k1\k1\k1$.  The 3 ghost-like NOT
gates in noisy-looking boxes indicate the possible errors.  At most
one of them is realized as an actual NOT gate $\Xc$.  It is the job of
the rest of the circuit to determine which, if any, of the Qbits
suffered such an extraneous NOT, and to restore the initial 3-Qbit
code state $\al\k0\k0\k0 + \be\k1\k1\k1$ regardless of which (if any)
of the Qbits was so corrupted.

The next two cNOTs, that target the upper ancilla, together with the
measurement gate on the upper ancilla, act as the parity-measuring
circuit of Figure 15, as do the final two cNOTs targeting the lower
ancilla and the other measurement gate.  Since the 3 Qbits in both basis
states for the code-word subspace have the same parity, the
measurement gate on the upper ancilla will indicate $x=1$ if and only
if either of the upper two code-word Qbits suffered a NOT error, while
the measurement gate on the lower ancilla will indicate $y=1$ if and
only if either of the lower two-code-word Qbits suffered such an
error.

Consequently if both measurements indicate 0, then there was no error, and
no correction to be applied.  If the upper measurement gate indicates
1 and the lower indicates 0, the upper code-word Qbit must have
suffered the error, and applying a NOT gate to it will restore the
original code-word state.  If the upper measurement gate indicates 0
and the lower indicates 1, the lower code-word Qbit must have suffered
the error, and the original code-word will be restored by applying the
NOT gate to the lower code-word Qbit.  And if both measurement gates
indicate 1 it is the middle code-word Qbit that suffered an error, and
to which the corrective NOT gate must be applied.  This is precisely
what the NOT gates on the right of Figure 19 accomplish.

How essential are the two measurement gates to this process?  Can the
corrective NOT gates be applied by judicious use of unitary controlled
gates, without having to perform any measurements at all?   Figure 20 
shows some steps in this direction.

\midinsert
\epsfxsize=7truein
\centerline{\epsfbox{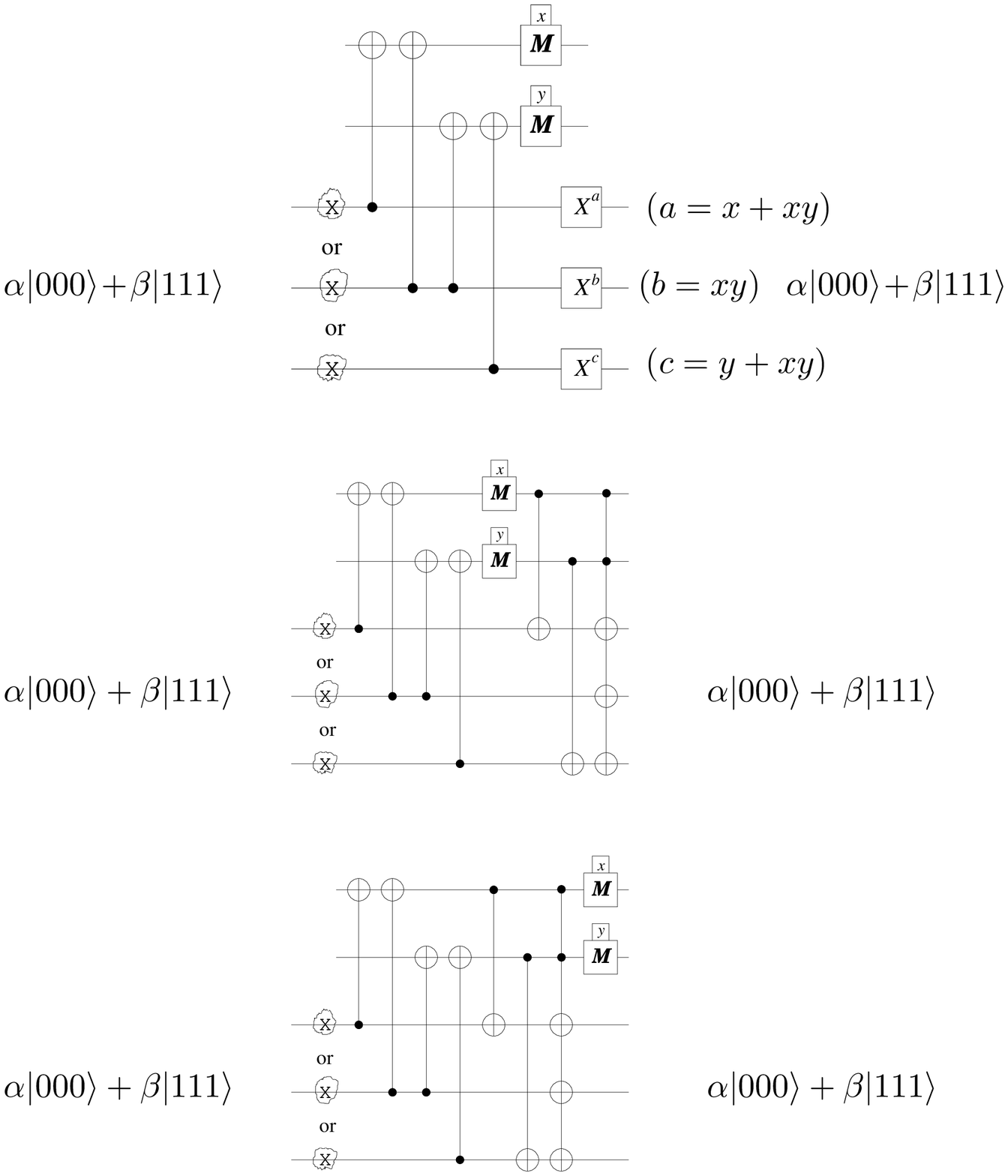}}
\vskip -60pt
\centerline{\bf Figure 20.}
\endinsert
\vfil\eject

\midinsert
\epsfxsize=7truein
\centerline{\epsfbox{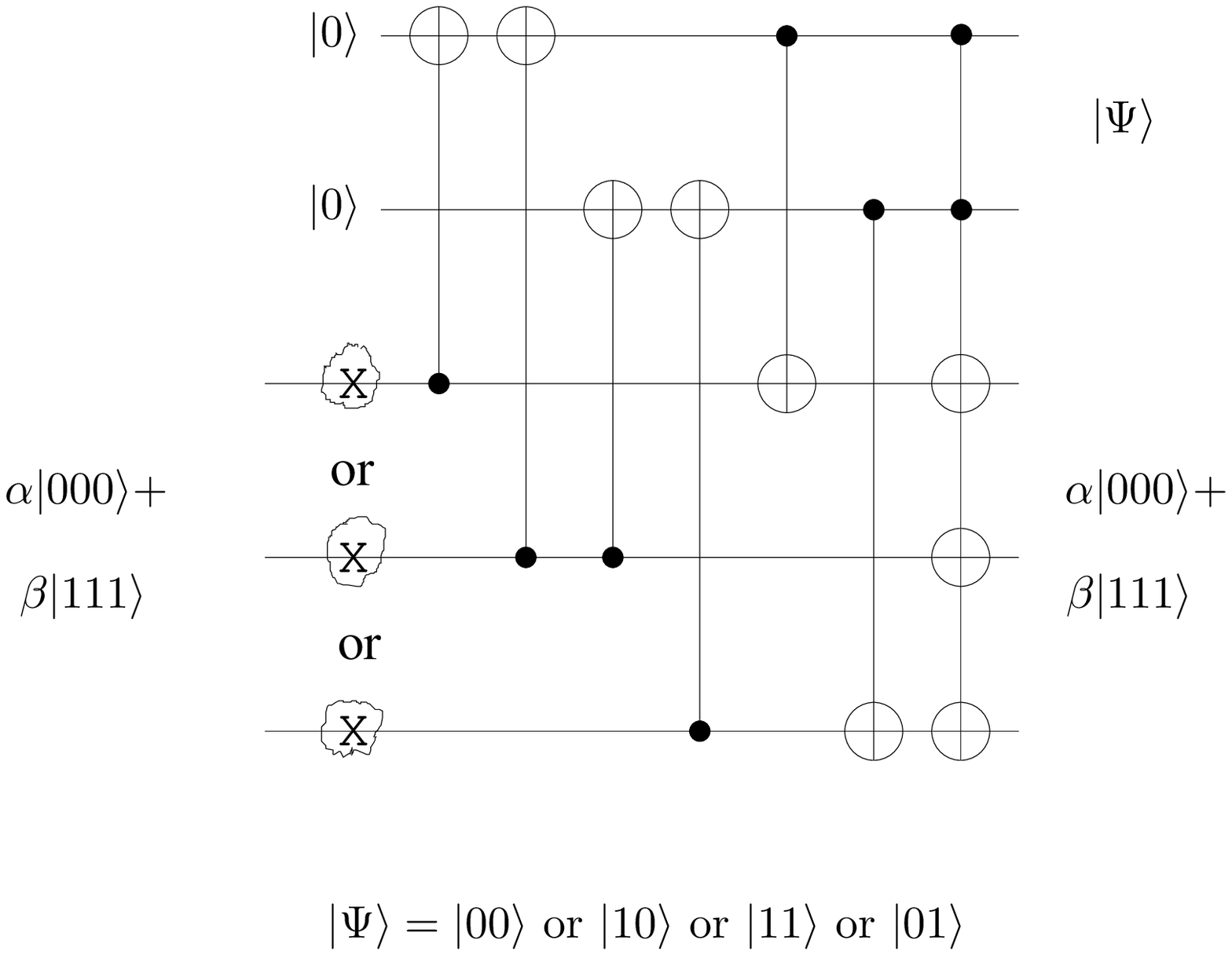}}
\vskip -240pt
\centerline{\bf Figure 21.}
\endinsert

The upper part of Figure 20 reproduces the relevant part of Figure
19.   The middle part replaces the action of the three corrective
NOT gates, contingent on the three measurement outcomes, by a
collection of unitary gates --- two cNOT gates and a doubly-controlled
triple-NOT gate (that can be built out of three Toffoli gates) ---
that are controlled by the post-measurement ancillary Qbits, and are
easily confirmed to apply just the required corrections.  

Now, just as in Figure 13, we may let these controlled gates
act prior to the measurement gates acting on their Qbits, without
altering either the outcomes or the probabilities of those outcomes.   
This is pictured in the lower part of Figure 20.   

But now the measurement gates act on the two ancillary Qbits only
after the error has been fully diagnosed and corrected by unitary
gates.  If we drop them from the circuit, as pictured in Figure 21,
the error has been corrected without using any measurement gates at
all.

The only problem with this circuit is the one we faced in trying to
produce a Qbit in the state $\k0$, without using a measurement gate.
There the measurement gate was needed to restore the ancilla to the
state $\k0$ so it could be used again.  It is the same for error
correction.  Having removed the measurement gates from the ancillas,
we no longer know whether their states are $\k0\k0$ (corresponding to
no error) or $\k1\k0, \k0\k1$ or $\k1\k1$ (corresponding to an error
in the top, bottom, or middle code-word Qbit).  To get ancillary Qbits
for the next round of error correction we must either initialize two
fresh Qbits to the state $\k0\k0$ by means of measurement gates, or
measure the two we already have, to determine which if any must be
flipped to restore their initial state.  In either case, measurement
gates must be employed.

\vskip 20pt

So, in summary, measurement is essential at both ends of a quantum
computation, nor is it problematic in quantum computer science. It's problematic character is softened because the state assigned to the Qbits is changed
discontinuously by all gates --- not just by measurement gates ---
thereby eliminating the disconnect between ``continuous Schr\"odinger
evolution'' and ``discontinuous collapse''.  To be sure, there remains
the distinction between linear, invertible, unitary gates with output
state fully determined by the input, and nonlinear, irreversible,
measurement gates, with output state stochastically determined by the
input.  But the actions of both types of gates are fully defined, with
nothing left to the discretion of the theoretical physicist.

Modulo unitary gates, all the diverse things that go under the name of
``measurement'' reduce to (multiple copies of) a single elementary
1-Qbit gate.  If there remains anything problematic about measurement
in quantum computer science, it lies entirely in the straightforward
action of the simple circuit pictured at the bottom of Figure 1.  And
because all measurements must be constructed out of 1-qubit gates, the
complex network of interactions (unitary gates --- often cNOT gates or
multiply controlled NOT gates) hidden inside many superficially
elementary measurements is always made explicit.  Bell's question, ---
what exactly qualifies as a measurement --- has a precise and
unambiguous answer.  A circuit may qualify as a measurement if and
only if contains at least one 1-Qbit measurement gate.

In quantum computation, there {\it is\/} a {\it system\/} (one more of
Bell's bad words) --- the computationally relevant Qbits --- and there
{\it are\/} things outside of it --- the gates that act on those
Qbits, the readings of the measurement gates, and the users who read
those readings.  There is no point to a computer unless {\it users\/}
learn something from it.  There {\it must\/} be coupling to something
on the outside (users) if the computer is to perform its function.

Does this generalize beyond quantum computation?  Unlike a computer,
the universe is not made just for us.  But {theoretical physics\/} is
made just for us.  There is no point to quantum mechanics unless {\it
we\/} learn something from it.  There {\it must\/} be coupling to
something on the outside (us) if the quantum theory is to perform its
function.  This is what I take to be the meaning of the Bohr quotation
at the head of this paper: ``In our description of nature the purpose
is not to disclose the real essence of the phenomena but only to track
down, so far as it is possible, relations between the manifold aspects
of our experience.''  Our description of nature is theoretical physics
--- quantum mechanics, in particular.  It is not concerned with ``real
essences'' and whether or not that term has any meaning, it certainly
does not within the setting of quantum mechanics.  What does have
meaning and what science strives to do is to impose an orderly
structure on our perceptions.

Compare quantum computation: in the theory of quantum computation the
purpose is not to disclose the real essence of the Qbits, but only to
track down correlations --- brought about by the action of
intermediate gates --- between our initial and final readings of the
elementary 1-Qbit measurement gates.

The world is, of course, more complex than a quantum computation.  A
computation uses a finite set of Qbits. It has an unambiguous
beginning and end. There is always a world external to the
computation. If there were not an outside world, there would be no
point in doing the computation because there would be nobody or
nothing to take advantage of the output.

Nobody (well practically nobody) wants to view the entire universe as
one colossal quantum computer, sufficient unto itself.  We should
adopt a similar modesty of scope in our view of quantum mechanics.
Physics is a tool for relating some aspects of our experience to other
aspects.  Every application of physics can be cast in a form that
begins and ends with statements about experience.  This (and not the
existence of parallel computations in parallel universes) is the great
lesson that quantum computer science teaches us.

\vskip 10pt

{\it Acknowledgment.\/} I would like to thank Anton Zeilinger for
twenty years of delightful conversations, and much excellent
hospitality in Vienna.  And I am deeply indebted to the U.S.~National
Science Foundation, for supporting my education and research for 50
years, most recently under Grant No.~PHY 0098429.

\vskip 10pt

\centerline{\bf References}

\medskip

\noindent 1. J.~S.~Bell, Physics World, 33-40, August, 1990.

\noindent 2. Niels Bohr,  {\it Collected Works}, 
(North-Holland, Amsterdam, 1985) Vol.~6, p. 296.

\noindent 3.  C. H. Bennett, Scientific American {\bf 257}, 108-116 (1987).

\bye